\documentstyle[preprint,aps]{revtex}
\begin{document}
\draft
\title {\bf Specific heat in the integer quantum Hall effect: An exact
diagonalization approach}
\author {Sudhansu S. Mandal and Muktish Acharyya\footnote{Present
Address: Institute for Theoretical Physics,
University of Cologne,
50923 Cologne, Germany}}
\address {Department of Physics,
Indian Institute of Science, Bangalore-560012, India\\
and\\
Condensed Matter Theory Unit,
Jawaharlal Nehru Centre for \\
Advanced Scientific Research,
Jakkur, Bangalore-560064, India}

\maketitle
\begin{abstract}
We have studied the integer quantum Hall effect at finite temperatures
by diagonalizing a
single body tight binding model Hamiltonian including Aharonov-Bohm
phase. We have studied the
energy dependence of the
specific heat and the Hall conductivity at
a given temperature. The specific heat shows a sharp peak between two
consecutive Hall plateaus. At very low temperatures,
the numerical results of the temperature
variations of specific heat
(in the plateau region)
are in good agreement with the
analytical results.
\end{abstract}

\vspace {0.4 cm}
\pacs{PACS number(s): 73.40.Hm, 71.30.+h \\
Key Words: Integer quantum Hall effect; Specific heat; Tight binding
          model}

%\narrowtext
Quantum Hall effect continues to draw considerable interest since
its surprised discovery \cite{exp} in two dimensional electronic
systems at very low temperatures and
in the presence of very high magnetic fields.
In this phenomenon, the Hall conductivity gets quantized
at integer filling factors $\nu$ with the value $\nu e^2/h$ in high
accuracy and the diagonal conductivity vanishes at the same time.
A similar phenomenon also occurs at certain fractional filling
factors as well \cite{rev}.
In this paper, however, we are concerned only with
the integer quantum Hall effect (IQHE).

The scaling theory \cite{tvr} predicts that
all the states are localized in two
dimensions due to disorder potential which respects time reversal
symmetry. In the presence of strong magnetic field, perpendicular
to the plane of the system, the electronic energy spectrum of
the system shows a series of Landau bands. Since the magnetic field
destroys time reversal symmetry and suppresses back scattering,
some of the states appear to be extended at the center of each
Landau band. On the other hand, the states away from the center of
each band remain localized.
The transport properties (electronic conduction) of the system
do depend on the distribution of localized and extended states.
This distribution of localized and extended states gives rise to
IQHE. So long as the Fermi level lies in the region of localized
states, the Hall conductivity remains pinned to a value forming
a plateau. The critical transition
\cite{huk} from one plateau to another
occurs when Fermi level passes through the region of extended states.

On the other hand, the equilibrium properties (say specific heat)
of the system depend simply on the density of states \cite{gor,ssc}.
These
properties cannot distinguish the localized states from extended
states. This is indeed so, as it emerges from our systematic
numerical diagonalization study in a lattice model. Agreeably, the
separation of electronic specific heat from lattice specific heat
in quantum Hall systems, from the experimental data, are extremely
difficult.
However, the experiments \cite{gor,bayot}
are performed in multilayer systems to extract electronic
specific heat.

In this paper, we consider a single body (spinless) tight binding
model Hamiltonian \cite{tbm} with random impurities in each site in
the presence of a uniform magnetic field perpendicular to the
lattice plane. We have calculated the Hall conductivity $\sigma_H$
and the specific heat $C_v$ at a finite temperature $T$ in a finite
lattice. We show the formation of Hall plateau and the transition
between two plateaus of the Hall conductivity by studying the
energy variation of the Hall conductivity. When the Fermi level
lies between two Landau bands, $C_v$ is exponentially small since
in such a case the probability of low energy excitation is very
small. In such a region we find $C_v \sim (1/T^2)e^{-\Delta/T}$,
where $T$ is the temperature and $\Delta$ is a typical interband
excitation gap. This behavior of $C_v$ is consistent with the
analytical results obtained from the self consistent Born
approximation. On the other hand, when Fermi level lies within a
Landau band $C_v$ increases substantially and shows a sharp peak.
This peak position of $C_v$ lies between two consecutive Hall
plateaus. We show analytically that $C_v$ in this region is, in
fact, proportional to the density of states at Fermi energy. $C_v$
reflects the profile of the density of states. The width of the
Hall conductivity plateau is larger than that of $C_v$ plateau.
This is due to the fact that while the transport properties
distinguish between localized and extended states, the equilibrium
properties do not.

Single body two dimensional tight binding Hamiltonian
is given by
\begin{equation}
H = - \sum_{<ij>} \left (e^{ia_{ij}}C_i^{\dagger}C_j + h.c. \right)
+ \sum_i w_i C_i^{\dagger} C_i
\end{equation}
\noindent  where $C_i^{\dagger}$ and $C_i$ are
spinless Fermionic creation and
destruction operators respectively at lattice site $i$.
Here $<ij>$ refers to nearest
neighbour hopping, and the hopping parameter is taken to be unity. A
uniform magnetic field $B$ is applied such that it produces a uniform
magnetic flux $\phi = (hc/e)/M$ ($M$ integer) per plaquette. When a
particle hops from site $j$ to site $i$, it acquires a phase
$a_{ij} = \int_j^i \vec A \cdot \vec{dl}$, where the vector potential
in the Landau gauge is taken to be $\vec A = (0, Bx)$. For this choice
of gauge, an electron gets nonzero phase while hopping along $y$
direction only. The (diagonal) second term in the Hamiltonian
corresponds to random disorder potential at site $i$ with strength
$w_i$ varying from $-W/2$ to $W/2$. We have constructed the Hamiltonian
(1) by considering the periodic boundary condition.
As a result, the energy spectrum
becomes $M$ number of Landau bands. Each band contains $L^2/M$ states
in a square lattice of size $L\times L$.

We compute the  Hall conductivity $\sigma_H$
using the Kubo formula \cite{lutin}
\begin{equation}
\sigma_H = {{i e^2 \hbar} \over {L^2}} \sum_m \sum_{n \neq m}
{{(f_n-f_m) <m|v_x|n><n|v_y|m>} \over {(E_m - E_n)^2}} \; ,
\end{equation}
\noindent where
$E_n$ is the energy of $|n>$-th state, and
\begin{equation}
v_{\tau} = (i/\hbar) \sum_j \left(C^{\dagger}_{j+\tau}C_j
e^{i a_{j+\tau,j}} - C^{\dagger}_j C_{j+\tau} e^{-i a_{j+\tau,j}} \right)
\end{equation}
\noindent is the velocity operator \cite{kun} along direction $\tau
=(\hat x ~~{\rm or }~~\hat y)$.
And $f_n = 1/(\exp [(E_n - E_F)/k_BT] + 1)$ is the Fermi function at
temperature $T$ with Fermi energy $E_F$. The electronic specific heat
per unit area of the system can be
computed using the formula \cite{huang}:
\begin{equation}
C_v = {1 \over {L^2k_B T^2}} \sum_n (E_n - E_F)^2 f_n (1-f_n) \; .
\end{equation}

Figure 1 shows the energy variations of $\sigma_H$ and $C_v$.
These are computed using the formulae (2--4)
for a system of size $12 \times 12$,
strength of random potential $W = 0.5$, flux per plaquette
$\phi = (hc/e)/12$, and at small temperature $k_BT = 0.01$.
Three distinct IQHE plateaus, between four lowest Landau bands, at
$\sigma_H = \nu (e^2/h) ~~(\nu = 1, 2, 3)$ are shown for $E_F < 0$.
It is observed that
the higher plateaus are formed
at relatively lower values \cite{seng} compared
to corresponding integral multiple of $e^2/h$. The decrease in the
value of quantization is due to the relatively higher values of
impurity strength for the formation of the corresponding plateaus
\cite{seng}.
The transitions between plateaus occur at the centers of the Landau
bands. In these transition regions, $C_v$ shows sharp but finite
peaks with the width same as
the Landau band width. In between the Landau
bands, $C_v$ is exponentially small since the low energy excitations
are the least probable and the plateaus in $C_v$ are also formed.
The widths of plateaus in $\sigma_H$ are relatively higher than that
of the corresponding plateaus in $C_v$. This is because of the fact that
the transport properties (like $\sigma_H$) are sensitive to extended
states or localized states which are formed at the tail of the Landau
bands but the equilibrium properties (like $C_v$) are insensitive to
the nature of the states.
This is indeed the case since the expression (4) for the specific
heat does not depend on the characteristics (localized or
extended) of the states.

Figure 2 shows the temperature variation of the specific heat at two
different energies corresponding to two different plateaus
(see Fig.1),
keeping other parameters fixed. The best fitted form of $C_v$ is
found to be $C_v \sim (1/{T^2}) e^{-\Delta/T}$ which is consistent
also with the numerical diagonalization result
\cite{tapas} for fractional quantum Hall
states. Here $\Delta$ is a
typical excitation gap which we find to be consistent
with the Landau band gaps.
The values of $\Delta$ for these two curves are found to
be almost same while the prefactors are different.

We now analytically evaluate the specific heat at the regions of the
plateaus and the transition between the plateaus separately at low
temperatures by using self consistent Born approximation (SCBA) for the
impurity potential, and compare those with the numerical results.
The density of states of an IQHE system in
SCBA \cite{ando} is given by
\begin{equation}
\rho(\epsilon) = {1 \over {2\pi l^2}} {2 \over {\pi \Gamma^2}}
\sum_n [\Gamma^2 - (\epsilon - \epsilon_n)^2]^{1/2} \; ,
\end{equation}
\noindent which is a series of semicircular bands of radius $\Gamma$
at Landau levels with energy $\epsilon_n
= (n + {1 \over 2}) \hbar \omega_c$. Here
$\omega_c$ is the cyclotron frequency and $l$ is the magnetic length
of the system. We assume that the broadenning $2\Gamma$ of
each Landau level is very small compared to $\hbar \omega_c$, i.e.,
$\Gamma << \hbar \omega_c$. The electronic specific heat of the system
can be expressed as
\begin{equation}
C_v = {1 \over {k_B T^2}} \int_0^{\infty}
\rho(\epsilon) (\epsilon-E_F)^2 f(\epsilon) [1-f(\epsilon)]
d\epsilon \; ,
\end{equation}
\noindent where $f(\epsilon) = 1/(e^{[(\epsilon - E_F)/k_BT]}
+ 1)$ is the Fermi function.

\bigskip

\noindent {\it (i) Specific heat at the plateau regions:}
The specific heat can be expressed using eqns (5) and (6) as
\begin{equation}
C_v = {C \over {k_B T^2}} \sum_n \int_{\epsilon_n - \Gamma}^{\epsilon_n
+ \Gamma} (\epsilon - E_F)^2 \left [ \Gamma^2 - (\epsilon -\epsilon_n)^2
\right]^{1/2} f(\epsilon) [1 - f(\epsilon)] d\epsilon \; ,
\end{equation}
\noindent with the constant $C = 1/(\pi^2 l^2 \Gamma^2)$. We now
consider that the Fermi energy $E_F$ is in between two
consecutive Landau bands.
In this case an integer number of bands are filled. By Taylor's expansion
of eqn (7) in powers of $\Gamma/\hbar \omega_c$, we obtain
\begin{equation}
C_v \approx {{C\pi \Gamma^2} \over {k_B T^2}}
\left[\sum_n (\epsilon_n - E_F)^2
f(\epsilon_n) [1- f(\epsilon_n)] + {\cal O}\left(
 {{\Gamma} \over {\hbar \omega_c}}\right)
^2 \right] \; .
\end{equation}
\noindent At low temperatures ($\hbar \omega_c/k_BT >> 1$), we perturbatively
evaluate $C_v$ with the expansion parameter \cite{ssm}
$e^{-\hbar \omega_c/k_BT}$. We thus find
\begin{equation}
C_v(T) = {4 \over {\pi l^2 k_B}} \left [E_F^2 + {{\omega_c^2 N^2}
\over 4} - E_F \omega_c N \right ]
{1 \over {T^2}} e^{-\hbar \omega_c/2k_BT} \; ,
\end{equation}
\noindent where $N$ is the number of filled
Landau bands. Note that the specific
heat is activated through the gap between two Landau levels.
Interestingly,
this is the
same temperature dependent form of $C_v$ that we have found
from our numerical diagonalization study. The prefactor in Eqn (9)
depends on the value of Fermi energy and the number of filled
Landau bands. For this reason, we obtain two different curves for
two different Fermi energies in Fig. 2.

\bigskip
\bigskip

\noindent {\it (ii) Specific heat at the regions of transition between
two consecutive plateaus:} In these regions,
Fermi energy lies inside a Landau band.
The approximations made in
$(i)$ will not hold good in this case since closeby states above the
Fermi energy are available for low energy excitations. Using
Sommerfeld's  low temperature expansion \cite{lutin}
of the energy derivative of Fermi
function in eqn (6), we obtain
\begin{equation}
C_v = {{\pi^2} \over 3} k_B^2 T \rho (E_F) \; .
\end{equation}
\noindent Clearly, the specific heat in the transition regions follows
the profile of the density of states which has maxima at the centers of
the Landau bands. This is indeed so that we have also obtained this
type of profile
from our numerical study. However, the linear temperature dependence
that we have obtained in eqn (10) does not resemble with our numerical
calculations.
In our numerical study, we have obtained
a fitted form $C_v \sim (1/T^2)
e^{-\Delta/T}$  here
as well, with $\Delta$ much less than that in the case when
Fermi energy lies between two successive Landau bands.
We remark that this fitted
form, in our numerical study, is a consequence of finite size diagonalization
study since in such a case, there always remains
small but finite gap between the
states within a Landau band. In other words, the linear temperature
dependence of $C_v$ in eqn (10), obtained from continuum description
of the density of states (Eqn (5)) within a Landau band, cannot be
achieved by a finite size diagonalization study. However, we believe
that in $L \to \infty$ limit, the linear temperature dependence of
$C_v$ may hold good.

In summary, we have studied the finite temperature Hall conductivity
and the specific heat of an IQHE system by finite size diagonalization
of a tight binding model Hamiltonian including Aharonov-Bohm phase. The
energy variation of Hall conductivity and the specific heat have been
studied at low temperatures. The specific heat also shows plateau
together with the Hall conductivity, and gets
sharply peaked between two consecutive plateaus. We have also calculated
specific heat analytically in self consistent Born approximation. The
temperature variation of specific heat (in the plateau region) agrees
well with the numerical results.
\bigskip
\bigskip

Acknowledgment:
Financial assistance from Jawaharlal Nehru Centre for Advanced Scientific
Research, Bangalore, is gratefully acknowledged.

\bigskip
\bigskip
\bigskip

\centerline {\bf Figure Captions}
\bigskip

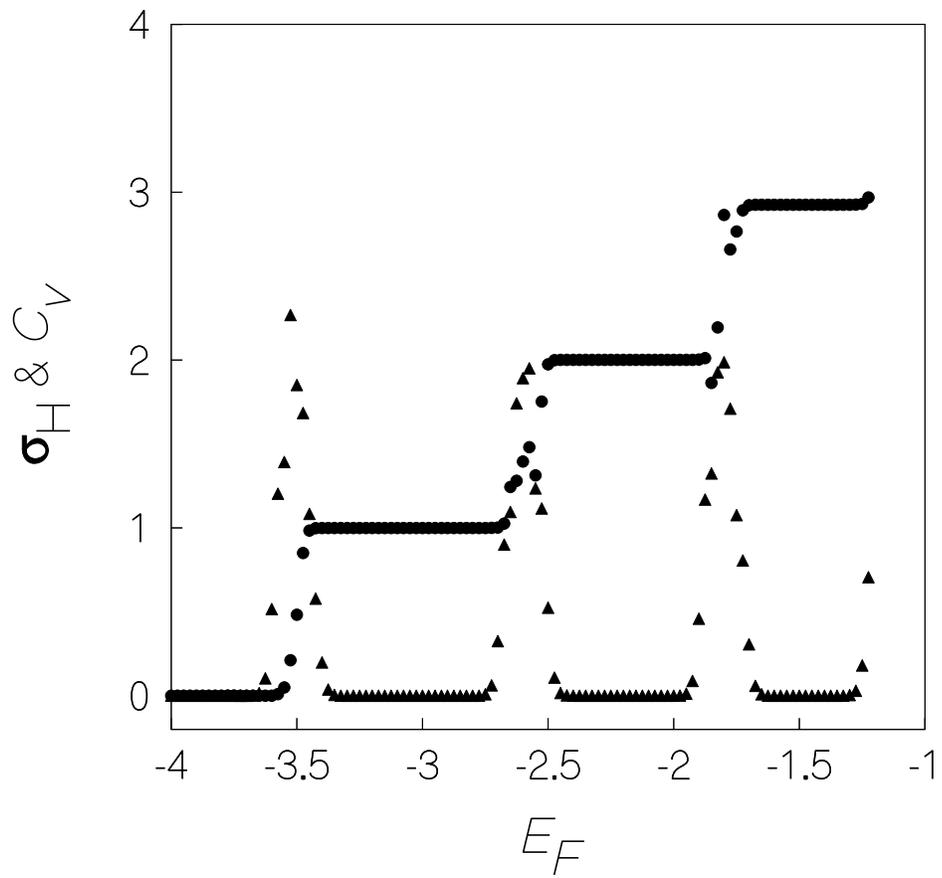
\noindent Fig. 1: Energy variations of $\sigma_H$ (circles) and $C_v$
(triangles), for $k_BT$ = 0.01, $W = 0.5$ and $M$ = 12. $\sigma_H$ is
shown in the unit $e^2/h$ while $C_v$ is in arbitrary unit.

\bigskip

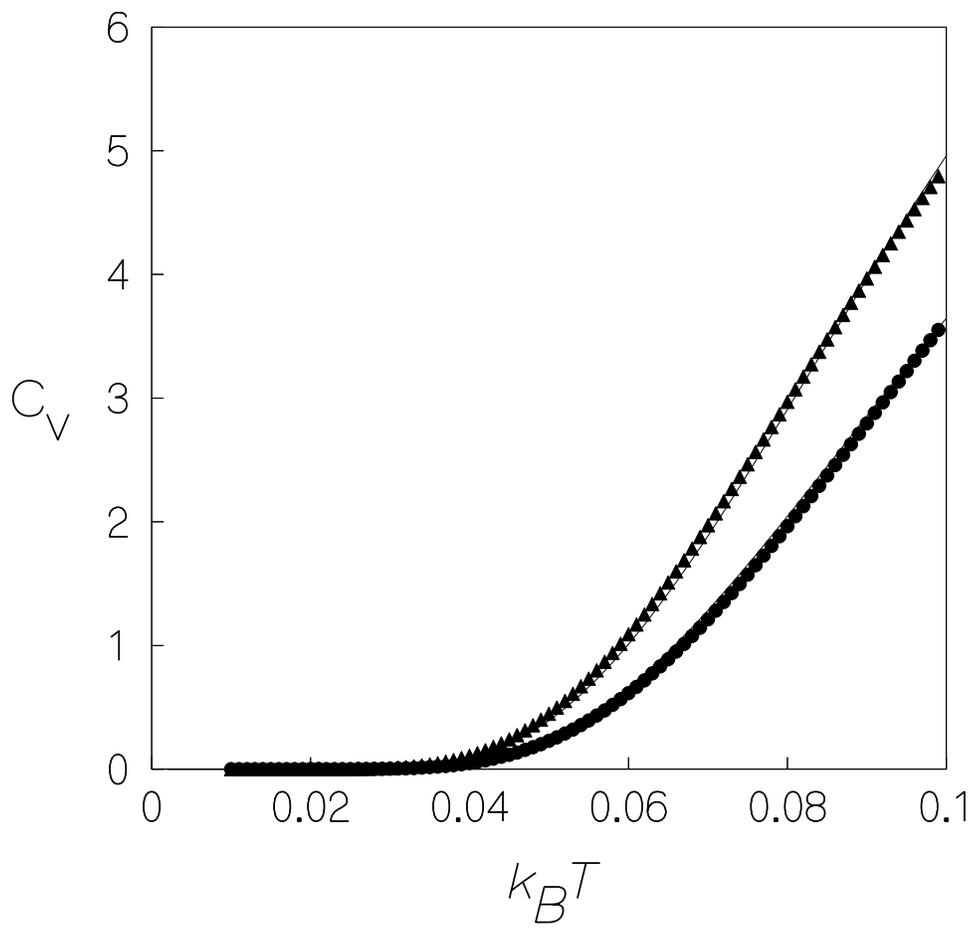
\noindent Fig. 2: Temperature variations of $C_v$ for two different
energies corresponding to two different plateaus.
(i) Circles represent $E_F=-3.0$ and (ii) Triangles represent
$E_F = -2.2$. Solid lines represent the best fitted  functional
form $C_v \sim (1/T^2) e^{-\Delta/T}$. Here $W = 0.5$ and $M = 12$.
Here, $C_v$ is shown in arbitrary unit.
\end{document}